\documentclass[]{an}

\usepackage[latin1]{inputenc} 
\usepackage[T1]{fontenc}
\usepackage[french,english]{babel}

\usepackage{amsfonts,amssymb}
\usepackage[fleqn]{amsmath}
\usepackage{graphicx}
\usepackage{pstricks}

\Pagespan{1}{}
\Yearsubmission{2012}

\def\Rm{\mathbb{R}}
\def\Zm{\mathbb{Z}}

\def\u{{\mathfrak u}}
\def\E{\vg{\cal E}}

\def\vg#1{\text{\mathversion{bold}$ #1 $}} 

\def\near#1{\lfloor #1\rceil }

\def\espace#1|{\noalign{\vspace{#1pt}}}
\def\gap#1{\vadjust{\vspace{#1}}}

\def \eqd{{\displaystyle{}
                   \mathop{=}^{\rm\scriptscriptstyle def}{}}}
\def\={{}:={}}

\def\Istd{{\bf I}}
\def\Bstd{\hbox{\bf $\textsf{B}$}}
\def\Qstd{\hbox{\bf $\textsf{Q}$}}
\def\Qstds{\hbox{\bf $\scriptstyle\textsf{Q}$}}
 
\def\Rstd{\hbox{\bf $\textsf{\small R}$}}
\def\Dstd{\hbox{\bf $\textsf{\small D}$}}
\def\Ustd{\hbox{\bf $\textsf{\small U}$}}

\def\Qred{\vg Q}

\def\Vstd{\hbox{\bf $\textsf{V}$}}

\def\defect{\delta_{\Qstds}\kern-0.5pt}   

\def\dd{{\rm d}}
\def\uu{{\rm u}}
\def\c{{\rm c}}

\def\ee{\hbox{\bf $\textsf{\small e}$}}
\def\vv{{\bf v}}

\def\vp{^{\vphantom{\rm T}}}

\def\log{\mathop{\rm ln}}

\def\sg{{\rm sg}}
\def\a{{\vphantom {b}{\rm a}}}
\def\b{{\rm b}} 
\def\d{{\rm d}} 

\renewcommand\r{{\rm r}}
\def\s{{\rm s}}

\def\Box{\raisebox{0.233mm}{\framebox[2mm][2mm]{}}}

\def\Boxdot{\raisebox{0.5mm}{\framebox[2mm][2mm]{}
\raisebox{-0.85mm}{$\kern-2.4mm \cdot $}}}

\def\arc{\mathop{\rm arc}\nolimits}

\def\G{{\cal G}}
\def\H{{\cal H}}

\renewcommand\appendix{\par  
\setcounter{section}{0}  
\renewcommand\thesection{Appendix \Alph{section}}  
}

\newcounter{remark}
\def\theremark{~\arabic{section}.\arabic{remark}}
\def\remark{\refstepcounter{remark}\textit{\theremark.}}

\newcounter{property}
\def\theproperty{~\arabic{section}.\arabic{property}}
\def\property{\refstepcounter{property}\textit{\theproperty.}}

\usepackage{times}
\begin{document}

\titlerunning{Integer-ambiguity resolution in astronomy and geodesy}
\title{Integer-ambiguity resolution in astronomy and geodesy}

\author{A. Lannes\inst{1,2}
\and
J.-L. Prieur\inst{3,4}\fnmsep\thanks{Corresponding author:
jean-louis.prieur@irap.omp.eu}
}
\institute{
CNRS/{\sc Supelec}/Univ Paris-Sud, France
\and
{\sc Supelec}, 2 rue E. Belin, 57070 Metz, France
\and
Universit\'e de Toulouse -- UPS-OMP -- IRAP, Toulouse, France
\and
CNRS -- IRAP, 14 avenue Edouard Belin, 31400 Toulouse, France
}

\received{2012 Oct 25} 


\abstract{%
Recent theoretical developments 
in astronomical aperture synthesis
have revealed the existence of integer-ambiguity problems.
Those problems,
which appear in the self-calibration procedures of radio imaging,
have been shown to be similar to 
the nearest-lattice point (NLP) problems
encountered in high-precision geodetic positioning,
and in global navigation satellite systems.
In this paper, 
we analyse  the theoretical aspects of the matter
and propose new methods for solving 
those NLP~problems.
The related optimization aspects concern
both the preconditioning stage, and the discrete-search stage
in which the integer ambiguities are finally fixed.
Our algorithms, which are described in an explicit manner,
can easily be implemented.
They lead to substantial gains in the processing time of both stages.
Their efficiency was shown via intensive numerical tests.
}

\maketitle

\section{Introduction}

Astronomical images obtained from ground-based observatories
are degraded by atmospheric turbulence. 
In particular,
the phase of the Fourier transform of the 
object-source distribution is severely perturbed
which leads to a significant loss of angular resolution 
in the resulting images. 
Thanks to the theoretical and technical developments 
of the last half century,
large interferometric arrays 
circumvent this difficulty in radio-astronomy, 
and now routinely provide sharp-edged images 
with a very high angular resolution.

One of the methods used for obtaining those nice results is
`self-calibration.'
In the most general case, the vectorial nature
of the electromagnetic field must be taken into account
in the very formulation of the problem;
see Hamaker (2000) and references therein.
In this paper,
we  however restrict ourselves to `scalar self-calibration.'
We thus refer to the same framework as that defined in 
Lannes \&~Prieur (2011).
In particular, we show that in the scalar case,
the phase-calibration problem
has a close similarity with the calibration problems encountered 
in high-precision geodetic positioning
and in global navigation satellite systems (GNSS). 

In fact, the approach we propose for solving the
phase-calibration problem in the scalar case 
is a good starting point
for tackling the more complex problem of full polarimetric phase calibration.
This possible extension however deserves a particular analysis which
goes beyond the scope of the present paper. 
Some guiding ideas for the corresponding 
`matrix self-calibration' approach 
are to be found 
in Hamaker (2000) and Yatawatta (2012).
The scalar case 
presented in Sect.~\ref{sec:2.1}
has already its own complexity. 
Any vectorial analysis should therefore
start from a good understanding
of that analysis.
We intend to address the global problem 
in a forthcoming paper.

In a previous paper 
(Lannes \&~Prieur, 2011), 
we have analysed the self-calibration procedure
in the scalar case. 
In that special case,
we have proposed a new approach to the problem:
the `arc-approach.'
The final step of that approach consists in solving 
a nearest-lattice-point (NLP) problem;
for a precise definition of this problem, see Sect.~\ref{sec:3.1}.

In fact, NLP problems appear in many fields 
of applied mathematics. 
In particular, as already mentioned, they play a central role
in high-precision geodetic positioning
and in~GNSS;
see, e.g., Lannes \&~Prieur~(2013).
In this paper, we present new methods for solving those NLP~problems.
These methods can therefore be applied both in astronomy and geodesy.

The standard way of solving an NLP problem
includes two stages:
a preconditioning stage, 
and a discrete-search stage in which 
the integer ambiguities are finally fixed.
The problem is usually preconditioned
by implementing the algorithm introduced by
Lenstra, Lenstra \& Lov\'asz~(1982): 
the LLL~algorithm. 
The {\sc Lambda} decorrelation method of Teunissen (1995)
can also be used for this purpose;
for the theoretical link between LLL-reduction 
and {\sc Lambda}-decorrelation,
see Lannes (2013).
The NLP~problem is then solved in the reduced basis 
thus obtained.
This is done via appropriate discrete-search techniques.
In this general context, 
we present our implementation of the LLL~algorithm,
as well as our discrete-search techniques.
This paper thus revisits and completes
the appendix~3 of Lannes \&~Prieur~(2011).
With regard to the current state of the art 
(see, e.g., Agrell et al.~(2002), Jazaeri et al.~(2012)),
the methods described in this paper lead to 
a speed-up of the order of two.

In Sect.~\ref{sec:2},
we show how those problems appear in astronomy and geodesy.
The main guidelines of our study are presented in Sect.~\ref{sec:3}.
Some basic notions are then defined
among which that of LLL-reduced basis.
In Sect.~\ref{sec:4}, we then describe an \hbox{LLL-type} algorithm allowing an
LLL-reduced basis to be built.
\hbox{Section~\ref{sec:5}} is devoted to the main contribution of our paper:
the discrete-search techniques to be implemented  
for finding the nearest lattice point
in the selected reduced basis.
We also describe
the techniques to be used
for identifying the points  
lying in some neighbourhood of the nearest lattice point.
Indeed, these points are also useful for the analysis of the related problems.
The computational issues of our contribution 
and its main results are summarized in Sects.~\ref{sec:6} and~\ref{sec:7}.

\vspace{-1mm}
\section{NLP problems in astronomy and geodesy}
\label{sec:2}

\vspace{-1mm}
We here present some NLP problems
encountered in astronomy (Sect.~\ref{sec:2.1})
and geodesy (Sect.~\ref{sec:2.2}).
The similarities
between the scalar case presented in Sect.~~\ref{sec:2.1},
and the global positioning problems of Sect.~\ref{sec:2.2}
are thus explicitly exhibited.

\vspace{-2mm}
\subsection{Self-calibration procedures in\\
\hspace{8mm}phase-closure imaging}
\label{sec:2.1}

When mapping incoherent sources
with aperture-synthesis devices, the pupil-phase perturbations
(hereafter pupil-phase biases) caused
by the atmospheric turbulence degrade 
the angular resolution of the restored image. 
A standard way for obtaining high angular-resolution images
is to estimate those pupil-phase biases from observations
of a calibrator (usually a reference star). 
However when the turbulence
is strong and quickly varies with time, 
this procedure is not possible. 
A way out is to use `self-calibration'
which corresponds to the situation where the object source
to be imaged plays the role of the calibrator.
Following the pioneering work of Cornwell \& Wilkinson (1981)
in the scalar case,
this problem can then be solved by alternate phase-calibration
operations and Fourier-synthesis processes. 
However, this procedure is
generally rather unstable. To ensure the reliability and 
the robustness of those techniques,
the phase-calibration operations must then be conducted with much care.

The model of the object source is refined throughout 
the iterative self-calibration procedure.
At each iteration,
the phase-calibration operation
consists in estimating
virtual pupil-phase biases~$ \alpha_\d(i)  $
so that the following equation is satisfied 
in a least-squares sense to be defined:
\begin{equation}
\exp \hbox{i}\varphi_\d(i,j)  \,
\exp -\hbox{i}[\alpha_\d (i) - \alpha_\d (j)]
=
\exp \hbox{i}\varphi_{\rm m }(i,j) 
\label{Eq:101}
\end{equation}
Here,
$ \exp \hbox{i}\varphi_{\rm d } $ 
and
$ \exp \hbox{i}\varphi_{\rm m } $ 
are the `phasors' of 
the (complex)  `visibility functions'
of the data and the model, respectively. 
The pairs $ (i, j) $, 
which define the edges of the `phase-calibration graph'~$ \G $,
correspond to the baselines of the interferometric device;
for further details, 
see Sect.~2 in Lannes \&~Prieur 2011.
The self-calibration procedure aims at 
reducing the phase discrepancy
\begin{equation}
  \varphi \eqd \varphi_{\rm d} - \varphi_{\rm m} 
\label{Eq:102}
\end{equation}
From Eq.~(\ref{Eq:101}),
we infer that the phase-calibration operation
consists in finding a function~$ \alpha _\d  $
such that the following relationship be valid
up to error terms: 
\begin{equation}
 \varphi(i,j) - [\alpha_\d(i) - \alpha_\d(j)] = 2\pi  N(i,j) 
\label{Eq:103}
\end{equation}
with $ N(i, j) $ in $ \Zm $.
In radio-astronomy, the related optimization problems are generally 
solved at the phasor level:
one minimizes the size of the chords 
associated with the phasors
\begin{equation*}
\exp \hbox{i}\{  \varphi(i,j) - [\alpha_\d(i) - \alpha_\d(j)] \} 
\end{equation*}
In some critical situations, the `chord functional' may have
several minima.
As shown in 
Lannes~(2005), 
and Lannes \&~Prieur (2011),
the analysis of the problem must then be conducted at the phase level.
We then consider the size of the quantities
\begin{equation*}
\arc \{\varphi(i,j) - [\alpha_\d(i) - \alpha_\d(j)] \}
\end{equation*}
where function $ \arc $ is defined as follows:
\begin{equation*}
\arc (\theta)  \eqd  
\theta - 2\pi  
\Bigl \lfloor {\theta  \over 2\pi} \Bigr\rceil \\
\end{equation*}
Here, $ \near x   $~denotes the integer of~$ \Zm $ closest to~$ x   $;
when $ x = k + 1/2 $ for some $ k $ in $\Zm $, $ \near x   $~is set equal to~$ k $.
The functional to be minimized is then of the form
\begin{equation}
g_\d (\alpha_\d ) \eqd
\| \arc(\varphi  -  B\alpha_\d) \| _w 
\label{Eq:104}
\end{equation}
where 
\begin{equation}
(B\alpha_\d)(i,j) \eqd \alpha_\d (i) - \alpha_\d (j)
\label{Eq:105}
\end{equation}
with $ \alpha_\d (1) = 0 $ for instance;
the norm $ \| \cdot \| _w  $ is defined as specified in Sect.~2.2 of
Lannes \&~Prieur 2011.

\vspace{1mm}

As explicitly shown in Sects.~4 to~7 of that paper,  
the arc approach gives a better insight into the problem.
The corresponding theoretical framework
appeals both
to algebraic graph theory (Biggs 1996)
and
algebraic number theory (Cohen 1996).
We now give a survey of the matter which shows how
those two main features are tightly imbricated.

The notion of  `phase closure,' 
which underlies the concept of `phase-closure imaging' (PCI),
is introduced in a context more general than that
usually defined in radio imaging and optical interferometry.
In particular, 
closure phases of order larger than three
may then be defined.
According to our algebraic-graph analysis,
the data-model discrepancy can be decomposed in the form
(see Sect.~3.3 of Lannes \&~Prieur 2011)
\begin{equation*}
\varphi  =  \varphi _\b + \varphi _\c 
\end{equation*}
Here, 
the baseline-bias function~$ \varphi _\b $ 
is equal to~$ B \alpha^{(\varphi )}   $ for\break

\vspace{-4mm}
\penalty-9900

\noindent
some~$ \alpha^{(\varphi)} $ (depending on~$ \varphi  $).
The function~$ \varphi _\c $
is the `closure function' of~$ \varphi  $;
it takes its values on 
the $ n_\c $~`closure edges' of~$ \G $,
the `loop-entry baselines' of the problem; 
see for example Figs.~3 and~4 in Lannes \&~Prieur (2011).

Clearly,
$ \varphi - B\alpha_\d    = \varphi _\c - B (\alpha_\d - \alpha^{(\varphi)})  $.
It then follows from
Eq.~(\ref{Eq:104})
that $ g_\d(\alpha _\d ) $ is equal to~$g(\alpha ) $ where
\begin{equation}
 g(\alpha) \eqd || \arc(\hat \varphi_\c - B \alpha) ||_w
\label{Eq:106}
\end{equation}
with
\begin{equation}
\alpha  \eqd \alpha_\d - \alpha^{(\varphi)}
\quad \hbox{and}\quad 
\hat \varphi_\c  \eqd \arc (\varphi_\c) 
\label{Eq:107}
\end{equation}
The minimizers of $ g_\d $ can therefore be 
easily deduced from those of~$ g $.

Divided by~$ 2\pi $, $ \hat \varphi_\c   $~defines 
some point~$ \vg{\hat \vv}$ of~$ \Rm^{n_\c} $.
We have shown that the minima of the arc functional~$ g $ 
are determined via particular `integer sets' 
associated with~$ \vg{\hat \vv}$.
Those integer sets correspond to some particular
points~$ \vg{\dot \vv}$ of lattice~$ \Zm^{n_\c} $;
see \hbox{Property~2} in Lannes \&~Prieur 2011.
In that algebraic-number framework,
finding the global minimizer of~$ g $ 
(and thereby that of~$ g_\d $) 
amounts to finding the point~$ \vg{\check \vv}$ 
of~$ \Zm^{n_\c} $ closest to~$ \vg{\hat \vv} $ 
with regard to some distance;
that distance is defined via some quadratic form
whose matrix~$ \Qstd $
is the inverse of the variance-covariance matrix~$ \Vstd $
of~$ \vg{\hat \vv}  $.
As explicitly clarified in Sect.~\ref{sec:3.1},
finding the global minimum
therefore amounts to solving a NLP~problem
in which $  \vg{\hat \vv} $~can be regarded as its `float solution.'
The main secondary minima of~$ g $, if any, 
correspond to $ \Zm^{n_\c} $-lattice points 
in some neighbourhood of~$ \vg{\check \vv}$.
Like for~$ \vg{\hat \vv}$, 
those points can be identified, in a systematic manner, 
via the integer-programming techniques presented in this paper.

\subsection{High-precision geodetic positioning}
\label{sec:2.2}

The techniques involved in
high-precision geodetic positioning
and global navigation satellite systems (GNSS)
are based on two types of data: 
the (carrier-)phase and code (or pseudo-range) observations; 
see, e.g., 
Lannes \&~\hbox{Gratton} (2009),
Lannes \&~\hbox{Teunissen} (2011).
The phase observational equations
of GNSS networks are of the form
\begin{equation}
\left| 
\begin{array}l
b_\kappa(i,j) - [\beta_{\r\kappa}(i) - \beta_{\s\kappa}(j)] = N(i,j) \\
\espace5|
\hbox{for $\kappa =1, \ldots, k$}
\end{array}
\right.
\label{Eq:108}
\end{equation}
In those problems, $ \kappa  $ is the epoch index;
$ k $~is the index of the current epoch;
$ \beta_{\r\kappa}(i) $ and~$ \beta_{\s\kappa}(j) $
are clock-phase biases.
Those biases, which are expressed in cycles,
depend on the frequency of the transmitted carrier wave;
subscripts r and s stand for receiver and satellite,%
\footnote{Here, satellite should be understood as satellite transmitter.}
 respectively;
$ i $~is the index of the receiver, and $ j $ that of 
the satellite;
$ N(i, j) $~is the integer ambiguity of the corresponding carrier-phase 
measurement.
The terms $ b_\kappa (i, j) $ include
the corresponding phase data
and 
the contributions associated with the real variables of the problem
other than the clock-phase biases:
position and atmospheric parameters, for instance;
see, e.g., de Jonge (1998) and Lannes \&~\hbox{Teunissen} 2011.
The set of receiver-satellite pairs~$ (i,j) $ 
involved in Eq.~(\ref{Eq:108})
forms the observational graph~$ \H_\kappa  $ 
of the GNSS~scenario of epoch~$ \kappa  $.
Owing to the particular structure of the phase equations~(\ref{Eq:108}),
the problem has a basic rank defect.
As outlied below, 
the latter can be eliminated by an appropriate redefinition of its variables.

In the system of Eqs.~(\ref{Eq:108}),
the GNSS functional~$ N $ takes its values
on~$ \G $,
the union of the graphs $ \H_\kappa  $ until the current epoch~$ k $.
The similarity of Eqs.~(\ref{Eq:103}) and~(\ref{Eq:108})
was first pointed out by Lannes \& Teunissen 2011.
In Lannes \& Prieur 2013, 
we were therefore led to propose for~$ N $
a decomposition
quite similar to that of~$ \varphi  $
in Sect.~\ref{sec:2.1}:
$ N = N_\b + N_\c $ 
with $ N_\b = B\mu^{(N)} $.
Here, $ \mu^{(N)} $ is
an integer-valued function depending on~$ N $;
$ \mu^{(N)} $~takes its values on the vertices of~$ \G $
other than the reference receiver~$ \r_1 $ (for example).
The corresponding `integer variable'
$ \vv \= N_\c $ 
lies in~$ \Zm^{n_\c} $
where $ n_\c $~is the number of closure edges of~$ \G $.
The redefined\gap{-1pt}
 clock-phase biases
are then of the form\gap{-1pt}
$ \beta_{\r\kappa}(i) + \mu_\r^{(N)}(i)  $ (for $ i\neq1$)
and 
$ \beta_{\s\kappa}(j) + \mu_\s^{(N)}(j)  $.

In a first stage, at each epoch $ k $,
the problem is solved in the LS~sense
by considering $ \vv $ as a `float variable.'
A~float solution~$ \vg{\hat \vv}$ is thus obtained and updated progressively.
In practice, this is done via recursive QR-factorization;
see Appendix C in Lannes \& Prieur 2013.
The ambiguity solution~$ \vg{\hat \vv}$
is then the point of~$ \Zm^{n_\c} $
closest to~$ \vg{\hat \vv}$ 
with regard to some distance. 
Like in PCI, 
that distance is defined via some quadratic form
whose matrix~$ \Qstd $
is the inverse of the variance-covariance matrix~$ \Vstd $
of the float solution~$ \vg{\hat \vv}  $.
In that case, the points of~$ \Zm^{n_\c} $
lying in some neighbourhood of~$ \vg{\check \vv}$
are involved in the corresponding validation techniques;
see \hbox{Verhagen} \&~\hbox{Teunissen}  2006.
Again, like in~PCI,
those points can be identified, in a systematic manner, 
via the integer-programming techniques presented in this paper.\penalty-9000
Once $ \vg{\check \vv}$ has been fixed and validated,
the real variables,
among which the redefined clock-phase biases,
are then estimated accordingly.

\vspace{-2pt}
\section{Guidelines}
\label{sec:3}

\vspace{-1pt}
This paper is essentially devoted to the methods 
to be used for solving
the NLP~problems encountered in astronomy and geodesy.
Setting $ n \= n_\c $,
we first define these problems as follows.

\vspace{-1pt}
\subsection{NLP problems}
\label{sec:3.1}

\vspace{-1pt}
Given some vector $ \vg{\hat \vv}$ of~$ \Rm^n $,
consider the (or a) vector~$ \vg{\check  \vv} $
of~$ \Zm^n $ such that
\begin{equation}
 \vg{\check  \vv} =
\mathop{\rm argmin}_{\vg \vv \in \Zm^n}
\| \vg \vv  - \vg{\hat \vv} \|^2_{\Qstds}
\label{Eq:1}
\end{equation}
The norm introduced here is that of~$ (\Rm^n, \Qstd) $:
the space~$ \Rm^n $
endowed with the inner product
\begin{equation}
  (\vv \mid  \vv')_{\Qstds} \eqd 
(\vv \cdot \Qstd \vv') 
\label{Eq:2}
\end{equation}
where $ \Qstd  $ is 
the inverse of
the variance-covariance matrix of
the `float solution'~$ \vg{\hat \vv}  $:
$ \Qstd \eqd \Vstd^{-1} $.
Clearly, $ (\;\cdot \;) $ is the
\hbox{Euclidean} inner product of $ \Rm^n $.
In matrix terms, we therefore have
\begin{equation}
(\vv \mid \vg{\vv'})_{\Qstds} 
= 
\vv^{\rm T} \Qstd \, \vg{\vv' } 
\label{Eq:3}
\end{equation}
All the quantities appearing in these equations
are expressed in the standard basis 
\begin{equation*}
\{ \ee_1, \, \ee_{\kern0.25pt 2}, \,\cdots,\, \ee_n \}
\end{equation*}
of~$ \Rm^n $ and $ \Zm^n $.
Note that this basis 
can be represented by the row matrix
\begin{equation}
\Bstd \eqd  
[ \ee_1 \quad   \ee_{\kern0.25pt 2} \quad  \cdots\quad   \ee_n]    
\label{Eq:4}
\end{equation}
whose entries are the vectors $ \ee_j $
for $ j = 1,\ldots, n $.

The integer lattice~$ \Zm^n $ regarded 
as a subset of~$ (\Rm^n, \Qstd) $
is denoted by~$ (\Zm^n, \Qstd) $;
$ \vg{\check \vv} $~is therefore 
a nearest lattice point to~$ \vg{\hat \vv} $ 
in~$ (\Zm^n, \Qstd) $.
Equation (\ref{Eq:1}) therefore defines an NLP~problem.

\subsection{Factorizations of~$ \Qstd  $}
\label{sec:3.2}

In this paper, we write the Cholesky factorization 
of~$ \Qstd  $ in the form
\begin{equation}
\Qstd = \Rstd^{\kern-0.5pt\rm T} \Rstd 
\label{Eq:5}
\end{equation}
where  $ \Rstd $ is an upper-triangular matrix.
Denoting
 \hbox{by~$ \|\cdot\| $} the Euclidean norm of $ \Rm^n $,
we therefore have, from Eqs.~(\ref{Eq:1}) and~(\ref{Eq:3}),
\begin{equation}
 \vg{\check  \vv} =
\mathop{\rm argmin}_{\vg \vv \in \Zm^n}
\|\Rstd(\vv  - \vg{\hat \vv}) \|^2
\label{Eq:6}
\end{equation}
Let $ \Dstd $ be the diagonal matrix defined via the relation
\begin{equation}
\Rstd = \Dstd^{1/2} \kern0.5pt \Ustd
\label{Eq:7}
\end{equation}
where $ \Ustd $ is an upper-triangular matrix 
whose diagonal elements~$ \uu_{j,j} $ are equal to unity.
For clarity, the diagonal entries of~$ \Dstd $ will be denoted 
by~$ \dd_j $.
From Eq.~(\ref{Eq:5}), we have
\begin{equation}
\Qstd = \Ustd^{\rm T} \kern-0.5pt\Dstd\, \Ustd
\label{Eq:8}
\end{equation}

\subsection{$ \Qstd $-Orthogonality defect}
\label{sec:3.3}

Any basis of~$ \Zm^n $ is characterized by a row matrix of the form
\begin{equation}
\vg B  \eqd  [ \vg e_1 \quad  \vg e_2 \quad  \cdots \quad  \vg e_n] 
\label{Eq:9}
\end{equation}
In general, such a basis
is far from being $ \Qstd $-orthogonal;
see Eq.~(\ref{Eq:2}).
To provide a measure of this defect,
we introduce the following notion.

\smallskip\noindent
{\sc Definition} 3.1. 
The parameter 
\begin{equation}
\defect(\vg B)\eqd 
\left(
{
 \prod_{j=1}^n   \vg e_j^{\rm T}  \kern-0.5pt 
\Qstd \kern0.5pt \vg e_{\kern-0.75pt j}\vp
\over
\hbox{det}\,\Qstd
}
\right)^{\kern-2pt 1/(2n) }
\label{Eq:10}
\end{equation}
is the `dilute $ \Qstd $-orthogonality defect' of~$ \vg B $~\Boxdot

\smallskip
\noindent
In the notation adopted in Eq.~(\ref{Eq:10}),
$ \vg e_j $~denotes the column matrix whose entries
are the components of the corresponding vector 
in the standard basis. 
Those entries therefore lie in~$ \Zm $.
Clearly, $ \hbox{det}\,\Qstd $~is the determinant of~$ \Qstd $.

According to Eqs.~(\ref{Eq:10}) and~(\ref{Eq:5}),
\begin{equation}
\defect(\vg B) =
\left(
{
 \prod_{j=1}^n   \| \vg b_j \|
\over
\hbox{det}\,\Rstd
}
\right)^{\kern-2pt 1/n }
\quad \quad 
 \vg b_j \eqd \Rstd \vg e_j 
\label{Eq:11}
\end{equation}
This relation shows that
$ \defect(\vg B) $~is the 
`dilute Euclidean-orthogonality defect' of the basis 
$ \{\vg b_1, \, \vg b_2, \,\cdots , \, \vg b_n \}   $.
It can be shown that $ \hbox{det}\,\Rstd $
is the volume of the $ n $-dimensional 
parallelepiped defined 
by these vectors.
Clearly,
$ \defect (\vg B) $~is greater than or equal to~1,
the zero defect corresponding to the case 
where $ \defect (\vg B)  = 1 $. 

\smallskip
The matrix  $ \vg M $ whose columns are the column matrices~$ \vg e_j $
of Eq.~(\ref{Eq:10})
is unimodular:
$ \vg M $  is an integer \hbox{$ n$-by-$ n $} matrix whose 
determinant is equal to~$ \pm 1 $.
The matrix relation
\begin{equation}
\vg B = \Bstd \vg M  
\label{Eq:12}
\end{equation}
gathers the vector relations
\begin{equation*}
 \vg e_j = \sum_{i=1}^n  m_{ij}\,\ee_i 
\quad \quad 
\hbox{(for $ j=1,\ldots, n $)} 
\end{equation*}
Clearly, the integers~$ m_{i,j} $ are the entries of~$ \vg M $.
In the same way as $ \vg M $ is associated with $ \vg B $,
the identity matrix~$ \Istd_n $ is associated with $ \Bstd $.
In terms of matrices,
we have
$ \vg e_j = \vg M\vg \ee_j  $, 
hence (from Eq.~(\ref{Eq:10}))
\begin{equation}
\defect(\vg B) =
\left(
{
 \prod_{j=1}^n   \vg \ee_j^{\rm T} \kern-1pt
\Qred \kern1pt\vg \ee_j\vp
\over
\hbox{det}\,\Qstd 
}
\right)^{\kern-2pt 1/(2n) }
\label{Eq:13}
\end{equation}
where
\begin{equation}
\Qred \eqd \vg M^{\rm T}\kern-1.pt\Qstd \vg M 
\label{Eq:14}
\end{equation}
Note that 
$ 
\hbox{det}\,\Qred 
=
\hbox{det}\,\Qstd 
=
(\hbox{det}\,\Rstd)^2
$.
To compute $ \defect(\vg B) $,
one is led to consider the factorization
\begin{equation}
 \Qred =
\vg U^{\rm T}\kern-2pt \vg D \kern0.5pt\vg U 
\label{Eq:15}
\end{equation}
where $ \vg U $ is an upper-triangular matrix 
whose diagonal elements~$ u_{j, \kern0.5pt j} $ are equal to unity;
$ \defect(\vg B) $
is then obtained via the logarithmic formula
\begin{equation}
\log\bigl( \defect(\vg B) \bigr) 
=
{1 \over 2n} 
\sum_{j=2}^n   \log
\Bigl(
1 +
 \sum_{i=1}^{j-1}   {d_i \over d_j}\vp u_{i,j}^2
\Bigr)
\label{Eq:16}
\end{equation}
where the~$ d_j $'s are the diagonal entries of~$ \vg D $.
Note that
\begin{equation}
\log\bigl( \defect(\Bstd) \bigr) 
=
{1 \over 2n} 
\sum_{j=2}^n   \log
\Bigl(
1 +
 \sum_{i=1}^{j-1}   {\dd_i \over \dd_j}\vp \uu_{i,j}^2 
\Bigr)
\label{Eq:17}
\end{equation}
As $ \Qred \eqd \vg M^{\rm T}\kern-1.pt\Qstd \vg M $
(Eq.~(\ref{Eq:14})),
$ \defect(\vg B) $ can also be regarded as 
the `reduction defect' of~$ \Qstd $ in basis~$ \vg B $, 
or in a 
more concise manner,
as the reduction defect of~$ \Qred  $.

In what follows, 
the guiding idea is to choose~$ \vg M $ so that
$ \defect(\Bstd) $~be reduced somehow:
\hbox{$ \defect(\vg B)  < \defect(\Bstd) $}.
The notion of reduced basis
introduced by Lenstra, Lenstra and Lov\'asz~(1982)
was a key step in that direction.

\subsection{LLL-reduced basis}
\label{sec:3.4}

{\sc Definition} 3.2. 
The column vectors $ \vg e_j $
of~$ \vg M  $  define an LLL-reduced basis of~$ (\Zm^n, \Qstd) $
if the matrix elements of~$ \vg U $ and $ \vg D $
in factorization~(\ref{Eq:15}) 
satisfy the conditions
\begin{equation}
|u_{i,j}| \le {1 \over 2}
\quad \hbox{for}\quad 
1 \le i < j \le n
\label{Eq:18}
\end{equation}
and
\begin{equation}
d_j  \ge (\omega - u_{j-1,\kern0.5pt j}^2) d_{j-1}
\quad \hbox{for}\quad 
2 \le j \le n
\label{Eq:19}
\end{equation}
with $ 1/4 < \omega < 1  $ \Boxdot

\medskip
Condition (\ref{Eq:18}) reduces $ \defect(\Bstd) $ 
by reducing the size of the matrix 
elements $ \uu_{i,j} $; see Eqs.~(\ref{Eq:17}) and~(\ref{Eq:16}). 
Condition~(\ref{Eq:19}) requires 
the $ d_j $'s be loosely sorted in increasing order
with no distinctive discontinuity;
the ratios~$ d_i / d_j $ (for $ i < j) $ are 
then made as small as `LLL$ _\omega  $-possible.'

\subsection{Statement of the NLP problem\\ 
\hspace{7mm} 
in the reduced basis}
\label{sec:3.5}

To complete Sect.~\ref{sec:3.1},
we now state the NLP~problem~(\ref{Eq:1})
in the selected reduced basis~$ \vg B $; 
see the context of Eq~(\ref{Eq:12}).
Clearly, 
$
\| \vv  - \vg{\hat \vv} \|^2_{\Qstds} 
=
\bigl \| 
 \vg{M}  \,
[\vg{M}^{-1}\kern-0.5pt( \vv  - \vg{\hat \vv}) ]
\bigr\|^2_{\Qstds} 
$.
Setting
\begin{equation}
\vg v \eqd \vg{M}^{-1} \vv
\quad \quad \quad 
\vg{\hat v} \eqd \vg{M}^{-1} \vg{\hat \vv}
\label{Eq:20}
\end{equation}
we therefore have
\begin{equation*}
\begin{array}l
\| \vv  - \vg{\hat \vv} \|^2_{\Qstds} 
=
\bigl \| 
 \vg{M} 
( \vg v  - \vg{\hat v}\kern0.5pt)
\bigr\|^2_{\Qstds} \\
\espace8|\kern14.2mm {}=
[ \vg v  - \vg{\hat v}\kern0.5pt] ^{\rm T}
\vg{M}^{\rm T}\kern-0.5pt
\Qstd
\vg{M}  
[ \vg v  - \vg{\hat v}\kern0.5pt]  \\
\end{array}
\end{equation*}
It then follows that
\begin{equation}
\| \vv  - \vg{\hat \vv} \|^2_{\Qstds} 
=
q(\vg v)
\label{Eq:21}
\end{equation}
where, from Eq.~(\ref{Eq:15}),
\begin{equation}
q(\vg v)
\eqd
\bigl \| 
\vg D^{1/2} \kern 0.5pt \vg U
\bigl( \vg v  - \vg{\hat v}\kern0.5pt\bigr) 
\bigr\|^2
\label{Eq:22}
\end{equation}
Let $  \vg{\check v} $ 
now be a vector of $ \Zm^{n} $ minimizing $ q(\vg v) $:
\begin{equation}
 \vg{\check v}  
=
\mathop{\rm argmin}_{\vg v \in \Zm^n}
q(\vg v)
\label{Eq:23}
\end{equation}
In the standard basis $ \Bstd $, 
the corresponding nearest lattice point 
is then obtained via the relation
(see Eq.~(\ref{Eq:20}))
\begin{equation}
\vg{\check \vv} 
=
 \vg M  \vg{\check v}
\label{Eq:24}
\end{equation}
To tackle the optimization problem (\ref{Eq:23}),
it is convenient to introduce the vector $ \vg{\tilde v} $ 
defined via the relation
\begin{equation}
\vg v -  \vg{\tilde v} 
\eqd 
\vg U\kern-1pt(\vg v -  \vg{\hat v}) 
\label{Eq:25}
\end{equation}
As the diagonal elements of $ \vg U $ are equal to unity,
the components of  $ \vg{\tilde v} $,
the `float conditioned ambiguities'~$ \tilde v_j  $,
are explicitly defined by the formula
\begin{equation}
\tilde v_j \eqd
\left|\kern -1pt
\begin{array}{ll}
\hat v_n& \hbox{if $ j=n $}\\
\espace5|
\hat v_j - \sum_{k=j+1}^n u_{j, k} (v_k - \hat v_k)
 & \hbox{if $ 1\le  j <n $}
\end{array}
\right.
\label{Eq:26}
\end{equation}
From Eqs. (\ref{Eq:22}) and (\ref{Eq:25}),
we have
\begin{equation}
 q(\vg v)
=
\sum_{j=1}^n  d_j (v_j - \tilde v_j)^2
\label{Eq:27}
\end{equation}
The discrete-search methods presented in Sect.~\ref{sec:5} 
derive from this equation.

\section{LLL reduction}
\label{sec:4}

In Sects.~\ref{sec:4.1} and~\ref{sec:4.2},
we introduce the reduction procedures that allow an LLL-reduced
basis to be built; see Sect.~\ref{sec:3.4}.
These procedures are basically involved in the LLL algorithm 
which provides all the related results.
Our version of this algorithm, which derives from that 
of Luo and Qiao~(2011),
is presented in Sect.~\ref{sec:4.3}.

Throughout this section, 
$ \vg D $ and~$ \vg U $ 
are the matrices of the factorization~(\ref{Eq:15}):
$ \Qred =
\vg U^{\rm T}\kern-2pt \vg D \kern0.5pt\vg U $
for
$ \Qred \eqd 
\vg M^{\rm T}\kern-1.pt\Qstd \vg M  $;
$ \vg M $~is some unimodular matrix.

\subsection{Procedure Reduce}
\label{sec:4.1}

If $ |u_{i,j}| > 1/2 $ for some  $ i < j $,
a procedure can be applied to ensure Condition (\ref{Eq:18}). 
This procedure is  referred to as {\sc Reduce}$ (i,j) $.

\medskip\noindent
{\bf Procedure R:} {\sc Reduce}$ (i,j) $

\smallskip\noindent
Consider the $ n $-by-$ n $ unimodular matrix
\begin{equation*}
\vg M_{\!i,j}\eqd \vg \Istd_n - \lfloor u_{i,j}\kern-1pt \rceil  \,
\vg \ee_i\vp  \vg \ee_j^{\rm T}
\quad \quad 
\hbox{($ i < j $)}
\end{equation*}
(Here, 
$ \vg \ee_i $~is the column matrix 
associated with the~$ i $th unit vector of~$ \Bstd $.)
Then, 
apply~$  \vg M_{\!i,j} $ to~$ \vg U $ and~$ \vg M $ 
from the right-hand side:
\begin{equation*}
\vg U \= \vg U\!  \vg M_{\!i,j}
\quad \quad \quad 
\vg M  \=\vg M \vg M_{\!i,j}
\quad \Boxdot 
\end{equation*}
Only the elements of the $ j $th~columns of $ \vg U $  and~$ \vg M $
can be affected by the action of~$ \vg M_{\!i,j} $:
$ u_{i' \!, j} \=  u_{i' \!, j} -    u_{i' \!, i}  \near{ u_{i,j}\kern-1pt} $ 
for all $ i' $, and likewise
$ m_{i' \!, j} \=  m_{i' \!, j} -   m_{i' \!, i}\near{ u_{i,j}\kern-1pt}$.\gap{-1pt}
Concerning~$ \vg U $,
as $ u_{i' \kern -1.2pt,\kern 1pt j} = 0 $ for $ i' > i $,
only the elements~$ u_{i' \kern -1.2pt,\kern 1pt j} $ 
for $ i' \le i $
are affected.  
In particular, $ u_{i,j} \= u_{i,j} -  \near{ u_{i,j}\kern-1pt} $.
In the updated version of~$ \vg U $, 
we thus have $ |u_{i,j}| \le 1/2 $.

\subsection{ Swap procedures}
\label{sec:4.2}

To ensure Condition~(\ref{Eq:19}), which is more subtle,
some particular procedure is to be implemented.
The core of the problem is then governed by
the $ 2 $-by-$ 2 $ matrices
\begin{equation}
D_j \eqd
\left[\kern-3pt
\begin{array}{cc}
d_{j-1} & 0\\
\espace3|
0 & d_j
\end{array}
\kern-3pt\right]
\label{Eq:28}
\end{equation}
and
\begin{equation}
U_j \eqd 
\left[\kern-3pt
\begin{array}{cc}
1 &\; u\\
\espace3|
0 &\; 1
\end{array}
\kern-3pt\right]
\kern 13.5mm
u \eqd u_{j-1, \kern0.5ptj}
\label{Eq:29}
\end{equation}
Setting (see procedure R)
\begin{equation}
M_j^{\rm r}   \eqd 
\left[\kern-3pt
\begin{array}{cc}
1 &\; - \lfloor u \rceil\\
\espace3|
0  &\; 1
\end{array}
\kern-3pt\right]
\label{Eq:30}
\end{equation}
we have
\begin{equation}
U_j\vp   M_j^{\rm r}   
=
\left[\kern-3pt
\begin{array}{cc}
1    &\; \breve u\\
\espace3|            
0   &\; 1
\end{array}
\kern-3pt\right]
\quad \quad \quad 
\breve u \eqd u -  \lfloor u \rceil
\label{Eq:31}
\end{equation}
Clearly, $  |\breve u|  $ is less than or equal to $ 1/2 $.

Now, consider Condition (\ref{Eq:19}) with
\hbox{$ u_{j-1, j} \= \breve u $}:
\begin{equation*}
d_j \ge (\omega - \breve u^2) d_{j-1}
\end{equation*}
When this condition is not satisfied, 
one is led to change the order of 
the corresponding ambiguity variables.
We then say that
\begin{equation}
M_j\vp \eqd M_j^{\rm r}\kern0.5pt S
\quad \quad \hbox{where}\quad \quad 
S \eqd
\left[\kern-3pt
\begin{array}{cc}
0 &\; 1\\
\espace3|
1 &\; 0
\end{array}
\kern-3pt\right]
\label{Eq:32}
\end{equation}
is a reduce-swap operator. 
From
Eqs.~(\ref{Eq:30}) and~(\ref{Eq:31}), 
it~follows that
\begin{equation}
M_j 
=
\left[\kern-3pt
\begin{array}{cc}
-\lfloor u \rceil &\; 1\\
\espace3|            
1 &\;    0
\end{array}
\kern-3pt\right]
\quad \quad  
U_j  M_j =
\left[\kern-3pt
\begin{array}{cc}
\breve u &\; 1\\
\espace3|            
1 &\;    0
\end{array}
\kern-3pt\right]
\label{Eq:33}
\end{equation}
Clearly,  $ U_j M_j $ is not an upper-triangular matrix.
Its original structure can be restored
as specified in the following property.
(The proof of this property is given in
\hbox{\ref{sec:A}.)}

\medskip\noindent
\textbf{Property RSR:} 
{\sc ReduceSwapRestore}

\medskip\noindent
Matrix
$ (U_j M_j)^{\rm T} D_j \,(U_j M_j)\vp  $
can be factorized in the form
\begin{equation*}
\bar U_j^{\rm T} \bar D_j \vp\kern0.8pt  \bar U_j\vp
\end{equation*}
where
\begin{equation*}
\bar D_j
\eqd
\left[\kern-3pt
\begin{array}{cc}
\bar d_{j-1}       &\; 0\\
\espace3|
0   &\;     \bar d_j\end{array}
\kern-3pt\right]
\kern10mm  
\bar U_j
\eqd
\left[\kern-3pt
\begin{array}{cc}
1 &\; \bar u\\
\espace3|
0 &\; 1
\end{array}
\kern-3pt\right]
\end{equation*}
in which
\begin{equation*}
\bar d_{j-1} \eqd
  d_j + \breve u^2 d_{j-1} 
\kern8mm \!
\bar d_j 
\eqd
d_j\,
{
d_{j-1}  
\over 
^{\vphantom {T^T}}\bar d_{j-1}
}
\kern8mm 
\bar u \eqd
\breve u \,
{
d_{j-1}   
\over 
^{\vphantom {T^T}}\bar d_{j-1}
}
\end{equation*}
As a corollary,
\begin{equation*}
G_j \kern0.5pt U_j  M_j
=
\bar U_j
\kern4mm
\hbox{where}
\kern4mm
G_j\eqd
\left[\kern-3pt
\begin{array}{cc}
\bar u&\quad  1- \breve u\bar u\\
\espace3|
1 &\quad   -\breve u 
\end{array}
\kern-3pt\right]
\end{equation*}
Moreover, $  [G_j^{-1}]^{\rm T}\kern-1pt D_j\vp G_j^{-1}  
= \bar D_j  $\quad \Boxdot

\medskip
The following procedure
in which $ u \eqd  u_{j-1, j} $ results from this property.

\medskip\noindent
{\bf Procedure  RSR:} {\sc ReduceSwapRestore}$ (j) $

\medskip\noindent
Compute $ \breve u = u -  \lfloor u\rceil  $, 
\begin{equation*}
\bar d_{j-1} = d_j + \breve u^2 d_{j-1} 
\kern8mm \!
\bar d_j 
=
d_j\,
{
d_{j-1}  
\over 
^{\vphantom {T^T}}\bar d_{j-1}
}
\kern8mm
\bar u
=
\breve u \,
{
d_{j-1}  
\over 
^{\vphantom {T^T}}\bar d_{j-1}
}
\end{equation*}
To update $ \vg D $,
set 
$ d_{j-1} \= \bar  d_{j-1} $
and
$ d_j \= \bar  d_j $.

\smallskip
Then, 
for $ j\ge 2 $, 
let
$
\vg M_{\!j} \eqd
\hbox{\bf diag}
([\kern0.75pt\vg \Istd_{j-2} \kern5pt 
M_j \kern5pt  \Istd_{n-j}] ) 
$
be the matrix obtained from the identity matrix~$ \Istd_n $
by substituting 
\begin{equation*}
M_j 
=
\left[\kern-3pt
\begin{array}{cc}
-\lfloor u \rceil &\; 1\\
\espace3|            
1 &\;    0
\end{array}
\kern-3pt\right]
\end{equation*}
for its $2$-by-$2$ block with largest diagonal index~$ j $;
see Eq.~(\ref{Eq:33}).
Likewise, define
$
\vg G_j \eqd
\hbox{\bf diag}
([\kern0.75pt\vg \Istd_{j-2}\kern5pt  G_j\kern7pt 
 \Istd_{n-j}] )
$
where
\begin{equation*}
G_j
=
\left[\kern-3pt
\begin{array}{cc}
\bar u&\quad  1- \breve u\bar u\\
\espace3|
1 &\quad   -\breve u 
\end{array}
\kern-3pt\right]
\end{equation*}
Matrices $ \vg U $ and $ \vg M $
are then updated as follows:
\begin{equation*}
\vg U \= 
\vg G_{\kern-0.75pt j}\vg U  \kern-1pt \vg M_{\!j}
\quad \quad \quad 
\vg M  \= 
\vg M  \kern-0.5pt \vg M_{\!j}
\quad \Boxdot
\end{equation*}

\smallskip
When implementing the operation 
$ \vg G_{\kern-0.75pt j} \vg U  \kern-1pt \vg M_{\!j} $,
the diagonal $ 2 $-by-$ 2 $ block of~$ \vg U $ 
with largest diagonal index~$ j $
is updated separately.
Indeed,
according to the corollary of \hbox{Property~RSR},
it~is equal to~$ \bar U_j $.

\smallskip
In the case where $  \lfloor u\rceil = 0 $, this procedure
reduces to \hbox{Procedure SR}:
{\sc SwapRestore}$ (j) $.

\subsection{LLL-type algorithms}
\label{sec:4.3}

The original LLL algorithm
provides the matrices~$ \vg U $ and~$ \vg   D $\gap{-0.5pt}
involved in the LLL-reduced version of~$ \Qstd $
(see Eqs.~(\ref{Eq:15}) and~(\ref{Eq:14})):
\begin{equation*}
\Qred  = 
\vg{U}^{\rm T} \kern-2.5pt  \vg{D} \kern0.5pt\vg{U}
\quad \hbox{for}\quad 
\Qred \eqd  
\vg M^{\rm T}\kern-1pt \Qstd \vg M 
\end{equation*}
It also yields
the  LLL-reduced basis~$ \vg B \eqd \Bstd \vg M $;
see Sects.~\ref{sec:3.3} and~\ref{sec:3.4}.
Its main instructions are the following
(see Eq.~(\ref{Eq:8}) for its initialization).

\bigskip
\noindent{\bf Original LLL algorithm}

\medskip

\hspace{-8.5mm}
\begin{tabular}{ll}
\vspace{0.5mm} 
1 &  $ \vg U \= \Ustd $; $ \vg D \= \Dstd $; 
      $ \vg M \= \vg \Istd_n $ \\
\vspace{1.5mm} 
2 & $ j \= 2 $\\
\vspace{1.2mm} 
\red\bf3 & {\red\bf while} $ j \le n $\\
\vspace{1.5mm} 
4 &\quad  if $ |u_{j-1, \kern 0.5pt j}| > 1/2 $, {\blue \sc Reduce}$ (j-1, j)$\\
\vspace{1.2mm} 
\blue 5 &\quad {\blue if} 
                $ d_j < (\omega - u_{j-1, \kern 0.5pt j}^2)\kern0.6pt  d_{j-1}  $\\
\vspace{1.2mm} 
6 &\quad \quad  
{\blue \sc SwapRestore}$ (j) $\\
\vspace{1.2mm} 
7 & \quad \quad $ j \= \hbox{max}(j-1, 2) $\\
\vspace{1.2mm} 
8 & \quad \blue else\\
\vspace{1.2mm} 
9 & \quad \quad for $ i \= j-2 $ down to 1\\
\vspace{1.2mm} 
10 & \quad \quad  \quad if $ | u_{i, j}| > 1/2 $, {\blue \sc Reduce}$(i,j)$\\
\vspace{1.2mm} 
11 & \quad \quad endfor 9\\
\vspace{1.2mm} 
12 & \quad \quad $ j \= j+1 $\\
\vspace{1.2mm} 
13 & \quad {\blue endif  5}\\
\red\bf 14 & {\red\bf endwhile 3}\\
\end{tabular}

\vspace{5mm}
Recently, 
Luo \& Qiao (2011) proposed a modified
LLL~algorithm
which can save a significant amount of operations, 
and also provides a basis for a parallel implementation.
In that approach,
which is justified via an example presented in
Sect.~3 of their paper,
the procedures imposing condition~(\ref{Eq:18})
are implemented at the end of this algorithm,
once the LLL condition~(\ref{Eq:19}) has been imposed.

\penalty-9900
\noindent{\bf LLL algorithm with delayed size-reduction}

\medskip

\hspace{-8.5mm}
\begin{tabular}{ll}
\vspace{0.5mm} 
1 &  $ \vg U \= \Ustd $; $ \vg D \= \Dstd $; 
      $ \vg M \= \vg \Istd_n $ \\
\vspace{1.5mm} 
2 & $ j \= 2 $\\
\vspace{1.mm} 
\red\bf3 & {\red\bf while} $ j \le n $ \hspace{9mm} 
                 [{\red to impose Condition (\ref{Eq:19})}]\\
\vspace{1mm} 
4 &\quad $ u \= u_{j-1, \kern 0.5pt j} $\\
\vspace{1mm} 
5 &\quad  if $ |u| > 1/2 $\\
\vspace{1mm} 
6 &\quad \quad $ {\rm ReduceOption} \= {\rm true} $\\
\vspace{1mm} 
7 &\quad \quad $ \breve u \= u - \lfloor u\rceil  $\\
\vspace{1mm} 
8 &\quad else \\
\vspace{1mm} 
9 &\quad \quad 
                $ {\rm ReduceOption} \= {\rm  false} $\\
\vspace{1mm} 
10 &\quad \quad $ \breve u \= u $\\
\vspace{1mm} 
11 &\quad endif 5\\
\vspace{1mm} 
\blue 12 &\quad {\blue if} 
                $ d_j < (\omega -\breve u^2)\kern0.6pt  d_{j-1}  $\\
\vspace{1mm} 
13 & \quad \quad 
               if  $ {\rm ReduceOption} = {\rm true} $ \\
\vspace{1mm} 
14 &\quad \quad \quad 
{\blue \sc ReduceSwapRestore}$ (j) $\\
\vspace{1mm} 
15 & \quad \quad else\\
\vspace{1mm} 
16 &\quad \quad \quad  
{\blue \sc SwapRestore}$ (j) $\\
\vspace{1mm} 
17 &\quad \quad endif 13\\
\vspace{1mm} 
18 & \quad \quad $ j \= \hbox{max}(j-1, 2) $\\
\vspace{1mm} 
19 & \quad \blue else\\
\vspace{1mm} 
20 & \quad \quad $ j \= j+1 $\\
\vspace{1.mm} 
\blue 21 & \quad \blue endif 12\\
\vspace{2.mm} 
\red\bf 22 & {\red\bf endwhile 3}\\
\red\bf 23 & {\red\bf for} $ j \= 2 : n $ \hspace{7mm}
             {[\red to impose Condition (\ref{Eq:18})}]\\
\vspace{0.8mm} 
24 &  \quad for $ i \= j-1 $ down to 1\\
\vspace{0.8mm} 
25 & \quad  \quad if $ | u_{i, j}| > 1/2 $\\
\vspace{0.8mm} 
26 & \quad \quad  \quad 
{\blue \sc Reduce}$(i,j)$\\
\vspace{0.8mm} 
27 & \quad \quad endif \\
\vspace{0.8mm} 
28 & \quad endfor 24\\
\vspace{0.8mm} 
\red\bf 29 &\red\bf  endfor 23\\
\end{tabular}

\bigskip
\noindent
Typically, this LLL algorithm 
with `delayed size-reduction' 
runs twice as fast as the original LLL algorithm.
Compared to the algorithm of Luo and Qiao (2011),
we made here the distinction between
the procedures RSR and~SR.
Some CPU~time can thus still be saved.
Those changes concern
the instruction blocks $ 5 $-$ 11 $
and \hbox{$  13 $-$ 17 $}.

The procedures described in 
Sects.~\ref{sec:4.1} and~\ref{sec:4.2}
can be completed so that this algorithm 
also provides the float solution in the LLL-reduced basis:
$ 
\vg {\hat v} = \vg M^{-1} \vg{\hat\vv}
$;
see Eq.~(\ref{Eq:20}).  
This can be done without 
forming $ \vg M^{-1} \kern-2pt$ explicitly.

\penalty-9900
According to Property RSR,
we have
\begin{equation*}
   \bar d_{j-1} = d_j  +\breve u^2  d_{j-1}  
\end{equation*}
Instruction~12 can therefore be equally 
well written in the form

\noindent
12\quad \quad \hspace{1.6mm}
if $ \bar d_{j-1} < \omega  d_{j-1} $

\smallskip\noindent
At level $ j $, 
the procedures RSR and SR modify, 
in particular, the matrix element~$ u_{j-2, \kern0.5pt  j-1} $.
As a result, this algorithm has a `one-step up-and-down structure;'
see instructions~18 and 20.
Lenstra, Lenstra and Lov\'asz have shown 
that for any~$ \omega  $ in the open interval
$ ]1/4\kern2mm  1[\kern1pt $, the algorithm terminates:
the number of times that the algorithm encounters the case where
$ \bar d_{j-1} < \omega  d_{j-1} $ is bounded.
In the limit case where $ \omega = 1 $, the convergence can also be guaranteed;
for further details, see Akhavi (2003),
Nguyen and Stehl\'e~(2009).

The convergence of the LLL algorithm is faster when reducing
the value of the relaxation parameter~$ \omega  $, 
but below some value (for example $ \omega = 0.70 $),
the dilute\penalty-9900
$ \Qstd $-orthogonality defect of the LLL-reduced basis~$ \vg B $ 
thus obtained begins to increase.
The choice of~$ \omega  $ therefore depends on the context.

For example, in GNSS, when handling a regional network in real-time
with $ n = 168 $ and $ \defect(\Bstd) \simeq  6.62 $,
$ \omega  $~may reasonably be set equal to~$ 0.9 $;
$ \defect(\vg B) $ can then be reduced to~$ 1.19 $ for example.
One then has a good compromise between 
the CPU~time required for finding the reduced basis,
and that used for the discrete search;
see Sect.~\ref{sec:5}.
On our old computers, the CPU time used for that LLL-reduction was
$ 0.075 $~second with our LLL-type algorithm,
against $ 0.141 $~second with the original LLL algorithm.
The LLL algorithm with delayed size-reduction 
effectively leads to
a gain of the order of two.

\looseness-1
For the statistical developments involved in the 
GNSS validation procedures, 
such as those of \hbox{Verhagen} and \hbox{Teunissen}  (2006),
the choice \hbox{$ \omega = 1 $} is preferable.
Indeed, as the discrete search is performed 
many times in the same reduced basis, 
the latter must 
be as $ \Qstd $-orthogonal as possible.

\section{Discrete search}
\label{sec:5}

This section is essentially devoted to the solution of the NLP~problem
in the selected reduced basis; 
see Sects.~\ref{sec:3.1},~\ref{sec:3.4}, ~\ref{sec:3.5}, and~\ref{sec:4.3}.
The problem is therefore to minimize $ q(\vg v) $ for $ \vg v $ lying in~$\Zm^n $;
see Eqs.~(\ref{Eq:23}) and~(\ref{Eq:27}).

Once the integer ambiguities 
$ v_n ,\, v_{n-1}, \ldots , \, v_{i+1}$
have been conditioned somehow (see the example given below),
Eq.~(\ref{Eq:26}) provides 
the float conditioned ambiguity
$ \tilde v_j $.

\smallskip\noindent
{\sl Example:  Babai point.}
Let us concentrate on Eq.~(\ref{Eq:27})
where the $ d_j $'s are 
loosely sorted in increasing order with no distinctive discontinuity.
To find a point $ \vg v $ for which $ q(\vg v) $ is a priori small, 
one is led to perform 
the `bootstrapping' recursive  process
described below. The point thus formed is   
the Babai point~$ \vg v^{\rm B} $ 
[Babai (1986)]:

\smallskip\noindent
Level $n$:

\noindent
$ v_n^{{\rm B}}  = \near{\tilde v_n }  \;$
where
$\; \tilde v_n = \hat v $ 

\smallskip\noindent
Level $n-1$:

\noindent
$ v_{n-1}^{{\rm B}}  = \near{\tilde v_{n-1}} \;$
where
$ \;\tilde v_{n-1} = \hat v_{n-1}  -  u_{n-1,n} (v_n^{{\rm B}} - \hat v_n\vp) $

\noindent
\hspace{3mm}$ \vdots $

\smallskip\noindent
Level 1:

\noindent
$ v_1^{{\rm B}}  = \near{\tilde v_1}  \;$
where 
$ \;\displaystyle
\tilde v_1 = \hat v_1 - \sum_{k=2}^n  u_{1,k} (v_k^{{\rm B}} - \hat v_k\vp) $

\penalty-9900
\noindent
The Babai point is often the solution of the NLP~problem, but not necessarily.
In any case however
(as explicitly shown in this section),
it is the `natural starting point' for searching this solution~\Boxdot

\subsection{Ambiguity conditioning
at level $ \vg j $}
\label{sec:5.1}

In the general case,
in the process of  conditioning ambiguity~$ v_j $,
we will use the following notation
(see Eq.~(\ref{Eq:27}))
\begin{equation}
 s_j \eqd
\sum_{i=j}^n  d_i (v_i - \tilde v_i)^2
\label{Eq:34}
\end{equation}
where $ \tilde v_i $ is given by  (see Eq.~(\ref{Eq:26}))
\begin{equation*}
\tilde v_i = 
\left|\kern -1pt
\begin{array}{ll}
\hat v_n& \hbox{if $ i=n $;}\\
\espace5|
\hat v_i - \sum_{k=i+1}^n u_{i , k } (v_k - \hat v_k)
 & \hbox{if $ 1\le  i <n $}
\end{array}
\right.
\label{Eq:35}
\end{equation*}
Note that
$
s_j =
 t_j  +  d_j(v_j - \tilde v_j)^2
$
where
\begin{equation}
t_j  \eqd
\left|\kern -1pt
\begin{array}{ll}
0&\hbox{if $ j=n $;}\\
\espace5|
\displaystyle
s_{j+1}
& \hbox{if $ j<n $}
\end{array}
\right.
\label{Eq:36}
\end{equation}
Let us now assume that the ambiguities $ v_n, v_{n-1}, \ldots, v_{i+1} $
have already been conditioned.
Denoting by $ \ell $ an integer candidate for $ v_j $, 
we then set
\begin{equation}
s \equiv s_j^{(\ell)} \eqd
t_j  +  d_j(\ell - \tilde v_j)^2
\label{Eq:37}
\end{equation}
The first ambiguity value~$ \ell $ to be considered at level $ j $ is then
\begin{equation}
 m =  \lfloor \tilde v_j\rceil  
\label{Eq:38}
\end{equation}
Indeed, $ |\ell - \tilde v_j| $ and thereby $ s $
are then as small as possible.
In the process of minimizing $ q(\vg v) $,
one is led to consider values of~$ \ell $ other than~$ m $.
These integers, 
$ \ell_1, \ell_2\,, \ldots, \ell_p\,, \ldots $,
where $ \ell_1 = m $,
are then sorted so that the discrepancies $ |\ell_p - \tilde v_j| $
form an increasing sequence.
The second integer to be considered is therefore~$ m+1 $ or~$ m-1 $.
Two cases are thus distinguished
(see Schnorr \& Euchner (1994)):

\smallskip\noindent
Schnorr$^{(+)} $:
$ m <  \tilde v_j $. 
Ambiguity $ v_j $ may then be conditioned at the successive terms of
the Schnorr list$^{(+)}$
\begin{equation*}
 m, \;m+1, \;m-1, \;m+2, \;m-2, \;m+3,\;\ldots 
\end{equation*}

\smallskip\noindent
Schnorr$^{(-)} $:
$ m \ge \tilde v_j $. 
Ambiguity $ v_j $ may then be conditioned at the successive terms of
the Schnorr list$ ^{(-)} $
\begin{equation*}
 m, \;m-1, \;m+1, \;m-2, \;m+2,\; m-3,\;\ldots 
\end{equation*}
In our implementation of the related approach,
we save CPU~time 
in the computation of the successive values
of~$ (\ell_p  - \tilde v_j)^2 $.
When handling the ambiguities~$ \ell $, 
and $ \ell  + 1 $ or $ \ell  - 1 $,
the following `perturbation formulas' are then used:
\begin{equation}
\left|\kern -1pt
\begin{array}l
[(\ell  + 1) - \tilde v_j]^2 
=
w^2 + (1 + 2w)\\
\espace7|
[(\ell  - 1) - \tilde v_j]^2 
=
w^2 + (1 - 2w)
\end{array}
\right.
\kern 3mm
w \eqd \ell - \tilde v_j
\label{Eq:39}
\end{equation}
The multiplication $ w^2 \= w \times w $ is then performed only for
$ \ell\= m $; see Sect~\ref{sec:5.2}.
Many multiplications can thus be avoided.
Note that the calculation of~$ 2w $ is then to be made in an
optimal manner
($ 2w $~is not necessarily computed as the sum $ w + w $).

\looseness-1
In the implementation of our approach,
we used object-orientated programming (OOP), 
and introduced a specific object referred to as {\blue SL}
(for Schnorr list).
More precisely, at the beginning of our program, we instantiated
an array of $ n $~such objects, one at each level~$ j $.
We then added two `methods' linked to this object: 
{\blue\sc Init} and {\blue\sc Next}. 
The latter are described
in the following section.

\subsection{Methods
{\blue I{\small NIT}} 
and 
{\blue {N\small EXT}}}
\label{sec:5.2}

The actions of 
{\blue\sc Init} and {\blue\sc Next}
consist in initializing and updating a two-element FIFO vectorial queue
$ (\ell_\a, \ell_\b) $,  $(s_\a, s_\b) $
associated with the two-component vector
$ (\ell,s) $.
The table below shows the structure of queue ($ \ell_\a, \ell_\b $) in the case 
of the Schnorr list$^{(+)}$:

\vspace{-0mm}
\begin{tabular}{lcccccc}
\smallskip
                   &$ \sg $      & $ \ell_\a $       &$ \ell_\b $\\
 \smallskip
     After {\blue\sc Init}:
&   $ +1$      & $m$       & $m$\\
\smallskip
      After {\blue\sc Next}:
&   $ -1 $   & $m $      & $m+1$\\
\smallskip
     After {\blue\sc Next}:
&    $ +1$    & $m+1$   & $m-1$\\
\smallskip
      After {\blue\sc Next}:
&    $-1 $   & $m-1$   & $m+2$
\end{tabular}

\vspace{2pt}
\noindent
Just before the call to {\blue\sc Init}, 
$ \tilde v_j $~is computed on the grounds of
Eq.~(\ref{Eq:26}); see Remark\ref{rem:5.1} further on.

\medskip\noindent
{\bf Method \blue I{\small NIT}:} instruction
$ (\ell, s) \= {\blue {\rm SL}_j} $--{\blue\sc Init}$  (\tilde v_j, t_j)  $

\noindent
Set\gap{2pt}\\
\null\quad 
$ \ell \= \near{\tilde v_j}  $\gap{2pt}\\
\null\quad 
$ w \= \ell - \tilde v_j $\gap{2pt} \\
\null\quad 
$ s \=  t_j + d_j\vp w^2 $\gap{2pt} \\
\null\quad 
$ \ell_\a \= \ell_\b \= \ell  $\gap{2pt} \\
\null\quad 
$  s_\a \= s_\b \= s $

\noindent
if $ w < 0 $\gap{2pt}\\     
\null\quad 
set $ \sg \= (+1) $\gap{2pt}\\
else\gap{2pt}\\
\null\quad 
set $ \sg \= (-1) $

\medskip\noindent
{\bf Method \blue N{\small EXT}:} instruction
$ (\ell, s) \= {\blue {\rm SL}_j} $--{\blue\sc Next}

\noindent
Set\gap{2pt}\\
\null\quad 
$  w \= \ell_\a  - \tilde v_j $\gap{2pt} \\
\null\quad 
$ \ell \= \ell_\a + \sg  $\gap{2pt}\\
if $ \sg = 1$\gap{2pt}\\
\null\quad 
$   s \= s_\a  + d_j (1 + 2w) $ \gap{2pt}\\
else\gap{2pt}\\
\null\quad 
$   s \= s_\a  + d_j(1 - 2w) $ \gap{2pt}\\
Set\gap{2pt}\\
\null\quad 
$ \ell_\a   \= \ell_\b  $;  $\; \ell_\b  \= \ell  $\gap{2pt}\\
\null\quad 
$ s_\a \= s_\b  $; $\; s_\b \= s $ \gap{2pt}\\
\null\quad 
$ \sg  \=  (-\sg)  $

\medskip\noindent
\textit{Remark}\remark
\label{rem:5.1}
According to  Eq.~(\ref{Eq:26}),
the float conditioned ambiguity $ \tilde v_j $ is given by the formula
\begin{equation}
\tilde v_j = 
\left|\kern -1pt
\begin{array}{ll}
\hat v_n & \hbox{if $ j=n $}\\
\espace5|
\tilde u_{j, \kern 0.5pt j+1} & \hbox{if $ 1\le  j <n $}
\end{array}
\right.
\label{Eq:40}
\end{equation}
where
\begin{equation}
\tilde u_{j, k} 
\eqd
\hat v_j - \sum_{\kappa = k}^n u_{j , \kappa} (v_\kappa - \hat v_\kappa)
\label{Eq:41}
\end{equation}
Now, consider the general case when $ \tilde v_j $ is to be computed,
when it has already been computed, 
and when in the meanwhile, for some $ j_{\rm r}> j $,
the integer ambiguities
$
v_{j_{\rm r}+1}, \,   v_{j_{\rm r}+2}, \,\ldots , \, v_{n -1}, \, v_n 
$
have not changed.
In our conditioning process, 
to reduce the corresponding CPU~cost,
$  \tilde v_j  $~is then computed as follows
(see Eqs.~(\ref{Eq:40}) and~(\ref{Eq:41})):\gap{5pt}\\
\null\quad 
If  $ j_{\rm r}  = n $ \quad (even if $ \tilde v_j $ has not been computed yet)\\
\null\quad \quad 
    $ \u \= \hat v_j $\\
\null\quad 
else\\
\null\quad \quad 
    $ \u \= \tilde u_{j, \kern0.5pt  j_{\rm r} +1}$\gap{4pt}\\
\null\quad 
for  $k \= j_{\rm r}  $ down to $ k \= j+1 $\gap{3pt}\\
\null\quad \quad 
  $ \u \=  \u  -  u_{j,k} (v_k - \hat v_k) $ \gap{2pt}\\
\null\quad \quad 
$ \tilde u_{j,k}  \=  \u $\gap{3.5pt}\\
\null\quad 
endfor\gap{2pt}\\
\null\quad 
 $  \tilde v_j  \= \u $ 

\smallskip\noindent
An auxiliary upper-triangular matrix~$ \vg{\tilde U} $ is thus built and updated through the process.
For further details, see Sect.~\ref{sec:5.3} and Remark\ref{rem:5.3}~\Boxdot

\subsection{Discrete-search algorithms}
\label{sec:5.3}

On the grounds of the notions introduced in Sects.~\ref{sec:5.1} and~\ref{sec:5.2},
we have designed three discrete-search algorithms
referred to as DS, DNS and DSC:

\begin{itemize}
\item[1)]
algorithm DS yields a  nearest lattice point~$ \vg{\check v} $ 
and $ \check q \eqd q(\vg{\check v}) $;

\smallskip
\item[2)]
algorithm DNS provides the first $ {\rm ns} $ NLP solutions\\
$ \vg{\check v}_1 \equiv\vg{\check v} $,
$ \vg{\check v}_2 $, 
$ \ldots $, 
$ \vg{\check v}_{\rm ns} $
with
$ \check q \equiv\check q_1 \le \check q_2 \le \cdots \le \check q_{\rm ns}$;

\smallskip
\item[3)]
given some parameter $ c > 0 $,
algorithm DSC identifies all the points $ \vg v $ of $ \Zm^n $
contained in the ellipsoid
\begin{equation}
\E(c) \eqd
\{
\vg v\in \Rm^n
:
q(\vg v) \le c
\}
\label{Eq:42}
\end{equation}
Clearly, $ \E(c) $ is centred on the float solution $ \vg{\hat v} $;
$ c $ defines the size of this ellipsoid.

\end{itemize}

\smallskip\noindent
\textbf{Algorithm DS.}
The objective is to condition the integer ambiguities~$ v_j $
so that $ q(\vg v) $~is minimum. 
We first note that
from Eqs.~(\ref{Eq:27}) and (\ref{Eq:34}),
\begin{equation}
\begin{array}l
q(\vg v) = s_1\\
\espace5|
\kern6.5mm{}= 
r_j + s_j 
\end{array}
\label{Eq:43}
\end{equation}
where
\begin{equation}
r_j \eqd \sum_{i=1}^{j-1} d_i 
(v_i - \tilde v_i)^2
\label{Eq:44}
\end{equation}
As $ r_j $ is non-negative,
we therefore have: 

\smallskip\noindent
{\sl Property\property\label{prop:5.1} 
If $ s_j \ge a $ for some $ a> 0 $, then $ q(\vg v) = s_1 \ge a $.}

\medskip
We first form the Babai point, here $ \vg v \= \vg v^{\rm B} $;
see the bootstrapping stage {$ 2 $-$ 8 $} of
the algorithm displayed in 
the next page.
All the Schnorr lists from $ j\=n $ down to 
\hbox{$ j\= 1 $},
as well as $ \vg{\tilde U } $,
are thus initialized;
see \hbox{Remark\ref{rem:5.1}}
with 
\hbox{$  j_{\rm r} = n $}.
As the Babai point 
is the first NLP~candidate,
we then set 
\begin{equation*}
 \vg{\check v}\= \vg v,
\quad 
q(\vg{\check v}) \equiv \check q \= s_1
\end{equation*}

The NLP search starts from the Babai point, 
but in the opposite sense,
with a Boolean variable $ {\rm Forwards} $ equal to $ {\rm true} $.
We therefore move to level $ j = 2 $.
Indeed, 
if $ v_1 $~was set equal to the next integer
of $ \blue{\rm SL}_1 $,
$ q(\vg v) $~would then be greater than $ \check q $.

To understand the principle of the algorithm in the general case,
let us assume that we are at some level~$ j \ge 2 $
with  $ {\rm Forwards}= {\rm true}  $.
We then consider the integer~$ \ell $ 
provided by 
{\blue ${\rm SL}_j $--{\sc Next}};
this method also
yields~$ s $: the new value of~$ s_j $  that would be obtained
if~$ v_j $ was set equal to~$ \ell $.
Clearly, $ s $~is greater than the current value of~$ s_j $
(and this would be worse with the remaining terms of the Schnorr list at this level).
Two cases are then to be considered.

\smallskip\noindent
{\sl Case 1:} $ s\ge \check q $.\gap{-0.5pt}  
If we then set $ v_j \= \ell $, 
whatever the conditioning of the integer ambiguities~$ v_{j-1}, \ldots, v_1 $,
we would then have
$ s_1 \ge \check q $
from Property~\ref{prop:5.1}.
Furthermore, another 
{\blue\sc Next}-type instruction
would increase~$ s_j $.
In this case, we are therefore left to move forwards to level 
\hbox{$ j \= j+1 $}.

\smallskip\noindent
{\sl Case 2:} $ s < \check q $.
As there is still a hope of
reducing~$ s_1 $
by  conditioning $ v_{j-1}, \ldots, v_1 $ in an appropriate manner,
we then set
\begin{equation*}
(v_j, s_j) \= (\ell, s), \quad 
t_{j-1} \=  s_j ,\quad 
 {\rm Forwards} \= {\rm false}
\end{equation*}
and move backwards to level 
\hbox{$ j \= j-1 $};
$ \tilde v_j $~is then updated;
note that
$ ( \near{\tilde v_j} - \tilde v_j  )^2 $ may then be smaller than previously at that level.

When the algorithm moves forwards to level \hbox{$ j\= j+1 $},
{\blue ${\rm SL}_j $--{\sc Next}}
is then called.
When it moves backwards to level \hbox{$ j\= j-1 $},
a new Schnorr list is  initialized via 
\hbox{{\blue ${\rm SL}_j $--{\sc Init}}}.
In both cases,
the situation is then analysed to define what is to be done;
see Cases~1 and~2.

Via Case 2, the algorithm may progressively reach level $ j=1 $
(several times).
If $ s  $~is less than $ \check q $,
$ \vg{\check v}  $ and~$ \check q $ are then updated;
see instructions~$ 32 $ to~$ 36 $.

Via Case 1, the algorithm reaches level~$ n $, 
at least once.
When 
{\blue ${\rm SL}_n $--{\sc Next}}
yields an~$ s $  greater than or equal to~$ \check q $,
the algorithm then stops; see instructions~$ 14 $ to~$ 25 $.
We then have the following property
(see Eq.~(\ref{Eq:16})):

\medskip
\noindent
{\sl Property\property\label{prop:5.2} 
At the end of the algorithm,
no point of~$ \Zm^n $ 
lies in the interior of ellipsoid~$ \E(\check q) $;
$ \vg{\check v} $~is on its boundary.}

\eject
{\bf\hspace{2mm} Algorithm DS}

\begin{tabular}{ll}
\vspace{0.70mm} 
1 & $ t_n \= 0  $; $ j_{\rm r} \= n $\\
\vspace{0.70mm} 
\blue\bf 2 & {\blue\bf for} $ j\= n $ down to 
$ j\= 1 $\hspace{5.5mm}  [{\blue\bf Babai loop}]\\
\vspace{0.70mm} 
3 &\quad Compute $ \tilde v_j $\\
\vspace{0.70mm} 
4 &\quad 
$ (\ell, s) \= {\blue {\rm SL}_j} $--{\blue\sc Init}
                                                      $  (\tilde v_j, t_j)  $\\
\vspace{0.70mm} 
5 &\quad  $ (v_j, s_j) \= (\ell, s) $\\
\vspace{0.70mm} 
6 & \quad if $ j >1 $ set $ t_{j-1} \=  s_j $\\
\vspace{0.70mm} 
\blue\bf 7 & {\blue\bf endfor} \blue 2\\
\vspace{1.25mm} 
\blue\bf 8 & $ (\vg{\check v}, \check q) \= (\vg v, s_1) $
                             \hspace{17.5mm}[{\blue\bf Babai point}]\\
\vspace{0.6mm} 
9 & $ {\rm NLPfound} \= {\rm false} $\\
\vspace{0.6mm} 
10 & $ {\rm Forwards} \= {\rm true} $ \\
\vspace{0.6mm} 
11 & $ j_1\vp \= 1 $; $ j^\star_2 \= 1 $\\
\vspace{0.70mm} 
12 & $ j \= 1 $\\
\vspace{0.70mm} 
\red\bf  13 & {\red\bf while} 
                      $ {\rm NLPfound} = {\rm false} $ 
                                     \hspace{6mm}[{\red\bf NLP search}]\\
\vspace{0.70mm} 
\red 14 & \quad  {\red if} $ {\rm Forwards} = {\rm true} $
                     \hspace{6.3mm} [{\red move forwards}]\\
\vspace{0.70mm} 
15 & \quad \quad if $ j=n $ \\  
\vspace{0.70mm} 
16 & \quad \quad \quad $ {\rm NLPfound} \= {\rm true} $\\  
\vspace{0.70mm} 
17 & \quad \quad else\\  
\vspace{0.70mm} 
18 & \quad \quad \quad $ j \= j+1 $\\
\vspace{0.70mm} 
19 &\quad \quad \quad 
$ (\ell, s) \= {\blue {\rm SL}_j} $--{\blue\sc Next}\\
\vspace{0.70mm} 
20 &\quad \quad \quad if $ s < \check q  $\\
\vspace{0.70mm} 
21 &\quad \quad \quad \quad   
                         $ (v_j, s_j) \= (\ell, s) $; $ t_{j-1} \=  s_j $\\
\vspace{0.70mm} 
22 &\quad \quad \quad \quad 
                              $ {\rm Forwards} \= {\rm false} $\\
\vspace{0.70mm} 
23 &\quad \quad \quad \quad 
   $ j_2 \= j $;  $ j^\star_2 \= 
                          \hbox{\rm max}(j_2\vp,  j^\star_2)  $ \\
\vspace{0.70mm} 
24 &\quad \quad \quad endif 20\\
\vspace{0.70mm} 
25 & \quad \quad endif 15\\  
\vspace{0.70mm} 
\red 26 & \quad  {\red else} \hspace{25.8mm} 
                                             [{\red move backwards}]\\
\vspace{0.5mm} 
27 & \quad \quad $ j \= j-1 $\\
\vspace{0.25mm} 
28 &\quad \quad  
               if  $ j < j_1 $ set  $ j_{\rm r}\vp \= j^\star_2 $\\
\vspace{0.70mm} 
29 &\quad \quad  
                  else \hspace{5.9mm}  set  $ j_{\rm r}\= j_2 $\\
\vspace{0.70mm} 
30 &\quad \quad 
                              Compute $ \tilde v_j $\\
\vspace{0.70mm} 
31 &\quad \quad 
$ (\ell, s) \= {\blue {\rm SL}_j} $--{\blue\sc Init}
                            $  (\tilde v_j, t_j)  $\\
\vspace{0.70mm} 
\red 32 &\quad \quad {\red if} 
         $ j = 1 $\hspace{1.5mm} [{\red case $ j = 1 $}]\\
\vspace{0.70mm} 
33 &\quad \quad \quad if $ s < \check q  $\\
\vspace{0.70mm} 
34 &\quad \quad \quad \quad  
                                          $ (v_1, s_1) \= (\ell, s) $\\
\vspace{0.70mm} 
35 &\quad \quad \quad \quad  
             $ (\vg{\check v}, \check q) \= (\vg v, s_1) $
  \hspace{13mm}[{\red\bf new $ \vg{\check v} $}]\\
\vspace{0.70mm} 
36 &\quad \quad \quad endif 33\\
\vspace{0.70mm} 
37 &\quad \quad \quad 
                   $ {\rm Forwards} \= {\rm true} $\\
\vspace{0.70mm} 
38 &\quad \quad \quad 
                     $ j_1\vp \= 1 $; $ j^\star_2 \= 1 $\\
\vspace{0.70mm} 
\red 39 &\quad \quad {\red else} 
               \quad \quad  [{\red case $ j > 1 $}]\\
\vspace{0.70mm} 
40 &\quad \quad \quad if $ s < \check q  $\\
\vspace{0.7mm} 
41 &\quad \quad \quad \quad  
                     $ (v_j, s_j)  \= (\ell, s) $; $ t_{j-1} \=  s_j $\\
\vspace{0.70mm} 
42 &\quad \quad \quad else\\
\vspace{0.70mm} 
43 &\quad \quad \quad \quad 
                       $ {\rm Forwards} \= {\rm true} $\\
\vspace{0.70mm} 
44 &\quad \quad \quad \quad $ j_1 \= j  $\\
\vspace{0.70mm} 
45 & \quad \quad \quad  endif  40\\
\vspace{0.70mm} 
46 & \quad \quad  \red endif  32\\
\vspace{0.70mm} 
\red 47 & \quad  \red endif  14\\
\vspace{0.70mm} 
\red\bf  48 & {\red\bf endwhile} \red\bf 13\\
\end{tabular}
\eject

\medskip
\noindent\looseness-1
{\sl Proof.}
Let us assume that there exists some $ \vg v^\circ $
in $ \Zm^n $
such that $ q^\circ \eqd q(\vg v^\circ) < \check q $.
From Eqs.~(\ref{Eq:27}) and~(\ref{Eq:26}), 
\begin{equation*}
\begin{array}l
q^\circ =
d_n \vp(v^\circ_n  - \hat v_n\vp)^2
+
d_{n-1} \vp(v^\circ_{n-1}  - \tilde v^\circ_{n-1})^2\\
\espace7|
\kern50mm{}
+
\cdots
+
d_1 \vp(v^\circ_1  - \tilde v^\circ_1)^2
\end{array}
\end{equation*}
The quantities 
\begin{equation*}
 s^\circ_n \eqd  d_n \vp(v^\circ_n  - \hat v_n\vp)^2
\end{equation*}
and
\begin{equation*}
\begin{array}l
 s^\circ_j \eqd  
s^\circ_n 
+
d_{n-1} \vp(v^\circ_{n-1}  - \tilde v^\circ_{n-1})^2\\
\espace5|
\kern20mm{}
+
\cdots
+
d_j \vp(v^\circ_j  - \tilde v^\circ_j)^2
\end{array}
\kern5mm 
(1\le j < n)
\end{equation*}
are then less than $ \check q $.
The algorithm starts by setting~$ v_n $ equal to the first
term of the Schnorr list~$ {\blue {\rm SL}_n} $.
In the NLP~search, it~then comes back to level~$ n $
via instruction~18, at least once,
until $ v_n $~is conditioned at~$ v^\circ_n $;
indeed, 
$ s^\circ_n  $ is less than $ \check q $
(see instructions~19 to~21).
The algorithm then starts moving backwards (via instruction~22),
and reaches instruction~31 with \hbox{$ j=n-1 $}.
The Schnorr list~$ {\blue {\rm SL}_{n-1}} $
is then systematically explored,
with possible excursions at levels $ j < n-1 $,
and this until $ v_{n-1} $ 
is set equal to~$ v^\circ_{n-1} $,
since $ s^\circ_{n-1} < \check q $.
The algorithm then proceeds to level~\hbox{$ n-2 $}.
It~then behaves similarly for that level,
and so on, until level $ j=1 $ 
where $ v_1 $~is set equal to the first term of~$ {\blue {\rm SL}_1} $.
The condition $ s < \check q $ of instruction~33 then holds,
hence 
via instructions~34 and~35,
$ \check q \= s \leq  q^\circ $, which 
contradicts the initial assumption.~\Box

\medbreak\noindent
\textit{Remark}\remark
\label{rem:5.2}
The pathological situations where $ \vg{\check v} $
is not the only point of~$ \Zm^n  $ on the boundary of $ \E(\check q) $
can be detected at level of instruction~$ 33 $.
The integer-ambiguity solution $ \vg{\check v} $ cannot then be validated.
As it is presented, the algorithm selects as solution the first~$ \vg v $  for which
$ q(\vg v) = \check q $; the other ones (if any) are discarded.
A subsequent statistical analysis can be used to diagnose such pathological cases.
In practice, as expected, such situations never occur 
\Boxdot

\medskip\noindent
\textit{Remark}\remark
\label{rem:5.3}
In the NLP search (instructions 13 to 48), 
the integers~$ j_1  $ and $ j_2 $  
keep track of the successive levels~$ j $ at which
the value of the Boolean variable {\rm Forwards} changes.
Note that $ j^\star_2 $~is the current largest index~$ j_2 $ at which the algorithm 
started moving backwards.
According to instructions 11 and~38,
whenever \hbox{$ j=1 $}, 
$ j_1\vp $~and~$ j^\star_2 $
are set equal to~$ 1 $.
By computing the float conditioned  ambiguities
in the framework of
Remark\ref{rem:5.1}
in which
$ j_{\rm r} $~is defined (from $ j_1\vp $, $ j_2\vp $ and $ j^\star_2  $)
via instructions~28-29,
the global CPU~time of algorithm~DS
can be reduced by a factor of the order of two.
In this context, 
the following technical point is also to be mentioned.

First of all, at the beginning of algorithm~DS,\gap{1.5pt} 
the values  of $ \hat v_j $ are placed on the diagonal
of $ \vg{\tilde U} $:
\begin{equation*}
 \tilde u_{j,\kern 0.5pt j} \= \hat v_j 
\quad \hbox{ (for $ j \= 1,\ldots, n $)}
\end{equation*}
Instructions~5, 21 and~41 are then completed by setting
\begin{equation*}
 v^*_j \= \ell - \hat v_j 
\end{equation*}
The instructions 
$ \u \= \hat v_j $ 
and 
 \hbox{$ \u \=  \u  -  u_{j,k} (v_k - \hat v_k) $}
of Remark\ref{rem:5.1}
are then changed into
$  \u \= \tilde u_{j,j}$
and
$ \u \=  \u  -  u_{j,k} v^*_k  $,
respectively.
The input variables of the function 
that computes $ \tilde v_j $ are then
$ j $, $ j_{\rm r} $, $ n $, 
$ \vg U $, $ \vg{\tilde U}  $ and $  \vg v^* $~\Boxdot

\medskip\noindent
\textit{Remark}\remark
\label{rem:5.4}
At the beginning of the NLP search,
the size parameter~$ c $ of the search ellipsoid~(\ref{Eq:42})  is defined
by the value
of $ q(\vg v) $ for the Babai point.
When the latter is not the NLP~solution,
$ c \equiv \check q $~is reduced via instruction~$ 35 $
\Boxdot

\medskip\noindent
\textbf{Algorithm DNS.}
The process is similar to that of algorithm~DS;
but, once the Babai point has been formed,
instead of moving forwards to level $ j = 2 $,
{\blue ${\rm SL}_1 $--{\sc Next}}~is 
set in motion $ {\rm ns} - 1 $ times.
We thus get a `working set' 
including $ {\rm ns} $ `candidate optimal lattice points' $ \vg{\check v}[{\rm ns}]  $
together with their $ q $-values $ \check q[{\rm ns}] $.
The last $ q $-value thus obtained, 
which (by construction) is larger than the previous ones,
is denoted by~$ \check q_{\rm ns} $.
In algorithm DNS, 
$ \check q_{\rm ns} $~is going to play the same role as $ \check q $ in algorithm~DS.

We then move forwards to level $ j = 2 $;
{\blue ${\rm SL}_2 $--{\sc Next}}
then provides the next term $ \ell $ of the Schnorr list at level~$ 2 $
together with the value of~$ s $ for that~$ \ell $.
If $ s $ is less than~$ \check q_{\rm ns} $,  
we then set $ (v_2, s_2) \= (\ell, s) $, $ t_1 \=  s_2 $,
and move backwards to level~$ 1 $; 
{\blue ${\rm SL}_1 $--{\sc Init}}
then defines (via~$ \ell $) some lattice point~$ \vg v $
with its $ q $-value $ q(\vg v) \= s_1 \= s $.
If $ s $ is less than~$ \check q_{\rm ns} $,
as  $ \vg v $ does not lie in the current set~$ \vg{\check v}[{\rm ns}]  $,
$ s $ and $ \vg v $ have to be inserted at their right places
in the sets~$ \check q[{\rm ns}] $ 
and~$ \vg{\check v}[{\rm ns}]  $;
the previous $ \check q_{\rm ns} $ and $ \vg{\check v_{\rm ns}} $
are then removed.
Instruction 
{\blue ${\rm SL}_j $--{\sc Next}}
is then performed until 
$ s $~is larger than the current value of~$ \check q_{\rm ns} $.
After each of these instructions, 
$ \check q[{\rm ns}] $ and~$ \vg{\check v}[{\rm ns}]  $
are of course updated and sorted.
In any case, we then finally move forwards to level 
\hbox{$ j = 2 $};
\hbox{{\blue ${\rm SL}_2 $--{\sc Next}}}
is then performed, and so on. 
Clearly, the principle is the same.

\medskip\noindent\looseness-1
\textbf{Algorithm  DSC.}
The process is again similar to that of algorithm DS.
As all the points of ellipsoid $ \E(c) $ are to be identified,
the tests $ s < \check q $ 
(the instructions 20, 33 and~40 of algorithm~DS) 
are replaced by $ s < c $.
When 
\hbox{{\blue ${\rm SL}_j $--{\sc Next}}}
is called,
we move forwards to level $ j+1 $, only when the value of~$ s $ thus obtained
is larger than (or equal to)~$ c $;
see Eq.~(\ref{Eq:42}) and \hbox{Property~\ref{prop:5.1}}.
Otherwise, we set 
\hbox{$ (v_j, s_j) \= (\ell, s) $}, 
\hbox{$ t_{j-1} \=  s_j $} and move backwards: 
\hbox{$ j\=j-1 $};
then 
\hbox{\blue ${\rm SL}_j $--{\sc Init}},
and so on.
Instruction 35 of algorithm~DS is replaced by other instructions which depend on
what is to be done with the vector $ \vg v $ thus identified;
see, e.g., Verhagen and Teunissen (2006), Lannes and \hbox{Prieur (2011)}.

\section{On some computational issues}
\label{sec:6}

The serial algorithms presented 
in Sects.~\ref{sec:4.3} and~\ref{sec:5.3} 
were implemented in C++ programs, 
and tested on old PC's working with Windows XP and Linux
operating systems.
Intensive testing was performed with real data on a regional GNSS~network.
As already mentioned 
at the end of Sect.~\ref{sec:4.3},  
for $ n=168 $, the CPU~time for the execution of 
our LLL-type algorithm with $ \omega =0.9 $ was negligible: 
about $ 0.075 $~second.
Compared to the original LLL~algorithm, 
as implemented for instance by Agrell et al. (2002)
or
\hbox{Jazaeri} et al. (2012),
the gain was of the order of two.
In fact, the parallel approach begins to be of interest for~$ n $
larger than
(say)~$ 200 $; 
see the reduction-list implementation 
of Luo and Qiao (2011).

Concerning the discrete-search algorithms presented in this paper,
our method was compared to that of
Jazaeri et al.~(2012) 
which corresponds to the present state of the art for the discrete search.
Our statistical study
on \hbox{$ 3 \times 10^5 $} \hbox{Gaussian} 
\hbox{$ \vg{\hat v} $-samples}
was conducted for $ n=168 $
in the \hbox{LLL-reduced} basis obtained as already specified.
The \hbox{Gaussian} \hbox{$ \vg{\hat v} $-samples} 
were of mean~$ \vg 0 $ 
and variance-covariance 
matrix \hbox{$ \vg V = \Qred^{-1} $}in that basis.
For each sample,  
$ \vg{\check v}_1  \equiv  \vg{\check v} $
and
$ \vg{\check v}_2 $
were determined via our DNS~algorithm;
see Sect.~\ref{sec:5.3}. 
The CPU times for those discrete searches were
$ 236 $~seconds with the algorithm of 
\hbox{Jazaeri} et al. (2012),
and
$ 129 $~seconds with our DNS~algorithm.
This gain, 
which is of the order of two,
is essentially due to the way 
of computing the float conditioned  ambiguities;  
see \hbox{Remarks\ref{rem:5.1}} and~\ref{rem:5.3}.

With regard to the self-calibration problems presented 
in Sect.~\ref{sec:2.1}, the previous statistical study gives and idea of the efficiency 
of our methods for finding 
the global and secondary minima of the arc functional~$ g $;
see Sect~\ref{sec:2.1}.

For handling the Schnorr lists at best,
some object-oriented programming tools have been introduced; 
see Sect.~\ref{sec:5.2}.
Our discrete-search algorithms
were thereby written in an `almost-electronic form.' 
Shortly,
they were designed for DSP (digital signal processor) 
implementation at the `speed of light.'
In GNSS, for example, the integer ambiguities of regional networks 
can thus be fixed in real time.
Let us finally note that for large $ n $,
the only discrete-search operations that can be performed in a parallel manner 
are those associated with the successive terms of the Schnorr lists at levels~$ n $
and~$ 1 $.

\section{Conclusion}
\label{sec:7}

\vspace{-1mm}
In this paper, we presented new methods for solving
the nearest-lattice point (NLP) problems
arising in astronomy, geodesy and GNSS.
The main theoretical aspects of the matter were also analysed.
This contribution concerns  both 
the preconditioning stage,
and the discrete-search stage
in which the integer ambiguities are finally fixed.
We proposed several algorithms whose efficiency 
was shown via intensive numerical tests on GNSS data.
The same algorithms can be used in the astronomical self-calibration
procedures. The related NLP problems are indeed very similar.

Concerning the preconditioning stage,
we have shown that 
the LLL-type algorithms with delayed size-reduction 
lead to a gain of  the order of two relative to the standard LLL~algorithm.
We have particularly optimized the discrete-search (DS) algorithms.
Our DS~algorithms run also about twice as fast as
the state-of-the-art DS algorithms of 
\hbox{Jazaeri} et al.~(2012).
We have thus been able to perform intensive calculations
on large-size problems with our old computers.
This would be particularly interesting for
real-time data processing of 
world-wide global GNSS networks.
As explicitly shown in Lannes (2013),
parallel versions of our LLL-type algorithms 
could also be implemented
for those extreme cases.

In astronomy, our self-calibration approach could lead to a 
substantial gain in computing time
for large interferometric arrays.
Another important asset of our approach
is to propose a method for validating
the calibration solution.
For each phase-calibration operation,
we determine the global minimum of the arc functional
and the first secondary minima (if any);
see Sects.~\ref{sec:2.1} and \ref{sec:5.3} in this paper,
and Sect.~5 in Lannes \& Prieur 2011.
In  the case of multiple minima,
the relative discrepancy between
the values the global and secondary minima
provides a measure against which 
the reliability of the process
can be appreciated. 
This is an innovative approach
which could promote the use of the self-calibration procedures
in radio imaging.
In particular, the extension of our approach to matrix self-calibration 
is an interesting problem that we intend to address in a forthcoming paper.

\appendix

\vspace{2pt}
\section{Proof of Property RSR}
\label{sec:A}

\noindent
The proof of Property
\hbox{{\sc ReduceSwapRestore}}
can be obtained as follows.

From Eqs.~(\ref{Eq:33}) and (\ref{Eq:28}), we have
\begin{equation*}
(U_j M_j)^{\rm T} D_j \,(U_j M_j)\vp 
=
\left[\kern-3pt
\begin{array}{cc}
\breve u &\; 1\\
\espace3|            
1 &\;    0
\end{array}
\kern-3pt\right]
\left[\kern-3pt
\begin{array}{cc}
d_{j-1} &\; 0\\
\espace3|            
0 &\; d_j   
\end{array}
\kern-3pt\right]
\left[\kern-3pt
\begin{array}{cc}
\breve u &\; 1\\
\espace3|            
1 &\;    0
\end{array}
\kern-3pt\right]
\end{equation*}
i.e.,  explicitly,
\begin{equation*}
(U_j M_j)^{\rm T} D_j \,(U_j M_j)\vp=
\left[\kern-3pt
\begin{array}{cc}
\bar d_{j-1}    &\quad      d_{j-1} \breve u\\
\espace4|            
d_{j-1} \breve u   &\quad  d_{j-1}\\
\espace2|            
\end{array}
\kern-3pt\right]
\end{equation*}
Let us now factorize this matrix in the form
\begin{equation*}
 \begin{array}l
\espace3|
{\rm U}^{\rm T} {\rm D} \vp {\rm U}\vp
=
\left[\kern-3pt
\begin{array}{cc}
1 &\; 0\\
\espace3|            
\u &\; 1   
\end{array}
\kern-3pt\right]
\left[\kern-3pt
\begin{array}{cc}
c_{j-1} &\; 0\\
\espace3|            
0 &\; c_j   
\end{array}
\kern-3pt\right]
\left[\kern-3pt
\begin{array}{cc}
1 &\; \u\\
\espace3|            
0 &\; 1   
\end{array}
\kern-3pt\right]  \\
\espace7|
\kern10.3mm
{}=  
\left[\kern-3pt
\begin{array}{cc}
c_{j-1}  &\quad  c_{j-1} \u\\
\espace5|            
c_{j-1} \u &\quad  c_j + c_{j-1} \u^2    
\end{array}
\kern-3pt\right]
\end{array}
\end{equation*}
By identifying the corresponding terms, we have
\begin{equation*}
 c_{j-1} =  \bar d_{j-1} 
\kern6mm
c_{j-1} \u = d_{j-1} \breve u
\kern6mm
 c_j + c_{j-1} \u^2  = d_{j-1}
\end{equation*}
As a result, 
$ \u = \bar u $
and
$ c_j + \bar d_{j-1} \u^2  = d_{j-1} $,
hence
\begin{equation*}
\begin{array}l
\displaystyle
c_j 
=
d_{j-1} - \bar d_{j-1}
\breve u^2\kern0.5pt
{
d_{j-1}^2  
\over 
^{\vphantom {T^T}}\bar d_{j-1}^2
}\\
\espace8|
\displaystyle
\kern3.1mm {} =
d_{j-1}
\left(
1 - \breve u^2\kern0.5pt
{
d_{j-1}  
\over 
^{\vphantom {T^T}}\bar d_{j-1}
} 
\right)\\
\espace8|
\displaystyle
\kern3.1mm {} =
{
d_{j-1}  
\over 
^{\vphantom {T^T}}\bar d_{j-1}
} 
(\bar d_{j-1} - \breve u^2 d_{j-1}) \\
\espace8|
\displaystyle
\kern3.1mm {} =
{
d_{j-1}  
\over 
^{\vphantom {T^T}}\bar d_{j-1}
}
\,d_j\\ 
\espace8|
\displaystyle
\kern3.1mm {} =
\bar d_j
\end{array}
\end{equation*}
Consequently,
$
( U_j M_j)^{\rm T} D_j \, (U_j M_j)\vp
= 
\bar U_j^{\kern0.5pt\rm T} \bar D_j \vp\kern0.8pt \bar U_j\vp
$.

\vspace{1mm}

The corollary results from the fact that
(see Eq.~(\ref{Eq:33}))
\begin{equation*}
\bar U_j (U_j M_j)^{-1} 
=
\left[\kern-3pt
\begin{array}{cc}
1 &\; \bar u\\
\espace3|            
0 &\;   1
\end{array}
\kern-3pt\right]
\left[\kern-3pt
\begin{array}{cc}
0 &\; 1\\
\espace3|            
1 &\; -\breve u   
\end{array}
\kern-3pt\right]  
=
\left[\kern-3pt
\begin{array}{cc}
\bar u&\quad  1- \breve u\bar u\\
\espace3|
1 &\quad   -\breve u 
\end{array}
\kern-3pt\right]
\end{equation*}
i.e.,
$ \bar U_j (U_j M_j)^{-1} = G_j $,
hence
$ G_j  U_j  M_j = \bar U_j $.
We then have
\begin{equation*}
\begin{array}l
(U_j M_j)^{\rm T} D_j \,(U_j M_j)\vp\\
\espace8| \kern 1mm{}=
(G_j U_j M_j)^{\rm T} \,
(G_j^{-1})^{\rm T}\kern-1pt  D_j\vp \kern0.5pt\vp G_j^{-1} \,
(G_j U_j M_j)\vp\\
\espace8| \kern 1mm{}=
\bar U_j^{\rm T} \bar D_j \vp
\kern0.8pt \bar U_j\vp
\end{array}
\end{equation*}
hence 
$ (G_j^{-1})^{\rm T}\kern-1pt  D_j\vp \kern0.5pt\vp G_j^{-1}  
= \bar D_j\vp $\quad \Box

\end{document}